\documentstyle[12pt]{article}
\def\ba{\begin{eqnarray}}
\def\ea{\end{eqnarray}}
\def\lb{\label}
\def\be{\begin{equation}}
\def\ee{\end{equation}}

\begin{document}

\begin{center}
{\Large \bf Functional equations for transfer-matrix operators in
open Hecke chain models}\footnote{This work was supported by the grants
INTAS 03-51-3350 and RFBR 05-01-01086-a .}
\end{center}

%\vspace{.5cm}

\begin{center}
\large{A.P.\,Isaev}
\end{center}

\begin{center}
Bogoliubov Laboratory of Theoretical Physics,
JINR, \\
 141980, Dubna, Moscow region, Russia \\
E-mail: isaevap@theor.jinr.ru
\end{center}

%\vspace{1cm}

{\bf Abstract.} We consider integrable open chain models formulated in terms of generators of 
affine Hecke algebras. The hierarchy of commutative elements (which are analogs of the commutative 
transfer-matrices) are constructed by using the fusion procedure. These elements satisfy a set of 
functional relations which generalize functional relations among a family of transfer-matrices in 
solvable spin chain models of $U_q(gl(n|m))$ type.

\section{Introduction}
\setcounter{equation}0

In our recent paper \cite{IsOg1} the non-polynomial baxterized solutions of reflection equations 
associated to the affine Hecke and affine Birman-Murakami-Wenzl algebras have been found. From 
these solutions one can produce
(see \cite{IsOg1} for details) polynomial solutions
proposed in \cite{Levi}, \cite{MudK} for the Hecke algebra case.
Relations to integrable
chain models with nontrivial boundary conditions were also discussed
in \cite{IsOg1}.

In this paper we concentrate to the investigation of chain models formulated in terms of generators 
of the affine Hecke algebra. Let braid elements $\{\sigma_1, \dots , \sigma_M \}$ and affine 
operator $y_1$ be generators of the affine Hecke algebra $\hat{H}_{M+1}(q)$
(for the definition of $\hat{H}_{M+1}(q)$ see
next Section). In \cite{IsOg1} we have suggested
to consider an open chain model with the Hamiltonian
\be
\lb{intrH}
{\cal H}_M = \sum_{k=1}^M \sigma_k +  (q-q^{-1}) \, \frac{ \xi}{y_1  -\xi}  \;\; \in 
\hat{H}_{M+1}(q) \; ,
\ee
where $\xi$ is a parameter of the model. This parameter fixes the boundary conditions for the chain 
and results from the non-polynomial solution \cite{IsOg1}
$$
y_1(x) = \frac{y_1 - \xi \, x }{y_{1} - \xi \, x^{-1}} \; ,
$$
of the reflection equation (see eq. (\ref{reflH}) below; $x$ is a spectral parameter) associated to 
the affine Hecke algebra. The model 
(\ref{intrH}) generalizes the $XXZ$ open spin chain model and, as it was shown in \cite{IsOg1}, 
this model describes universally a set of integrable systems for any representation of the algebra 
$\hat{H}_{M+1}(q)$. The most interesting representations of $\hat{H}_{M+1}(q)$
(from the point of view of applications) are
the so-called $R$-matrix representations.
It is known that the Hecke algebra can be realized via
$R$-matrices in the fundamental representation for $U_q(gl(n))$ \cite{Jimb1}, \cite{FRT}
and $U_q(gl(n|m))$  \cite{Isa}. For $U_q(gl(n|m))$ case this realization is
$$
\rho(\sigma_i) = I^{\otimes (i-1)} \otimes \hat{R} \otimes I^{\otimes (M-i)} =: \hat{R}_{i i+1} \; ,
$$
where $I \in Mat(n+m)$ is the identity  matrix and $\hat{R}$ is the braid form of fundamental 
$R$-matrix for $U_q(gl(n|m))$ (the explicit form of $\hat{R}$ is presented in
Conclusion). For the affine element we have
$$
\rho(y_1)= L \otimes I^{\otimes M} =: L_1 \; ,
$$
where elements of $L \in Mat(n+m)$ are generators of $U_q(gl(n|m))$-type reflection equation
algebra
$\hat{R}_{12} L_1 \hat{R}_{12} L_1 = L_1 \hat{R}_{12} L_1 \hat{R}_{12}$.
All these $R$-matrix representations $\rho$ lead to integrable chain systems with the Hamiltonians 
$\rho({\cal H}_M)$, where ${\cal H}_M$ is defined in (\ref{intrH}). Considering the special case of the $U_q(gl(2))$-type representation $\rho$ 
of the algebra $\hat{H}_{M+1}$ we reproduce from (\ref{intrH}) a Hamiltonian for the $XXZ$ open 
spin chain model with the general boundary condition for one side of the chain. In view of this we 
call the model formulated in terms of the generators of 
$\hat{H}_{M+1}$ with the universal Hamiltonian (\ref{intrH}) as the {\it open Hecke chain model}.

In this paper the hierarchy of commutative elements
(which are analogs of the transfer-matrices) are constructed
for open Hecke chain model by using the algebraic version of the fusion procedure \cite{KRSk}, 
\cite{KuSk82}, \cite{MeNe}. The transfer-matrix type elements satisfy functional relations which 
generalize functional relations for transfer-matrices in solvable open chain models (see, e.g., 
\cite{Zho}, \cite{NZ} and refs. therein).

The paper is organized as follows. In Sect. 2 we give the definition of the affine Hecke algebra 
and recall the algebraic formulation \cite{IsOg1} of the system of open Hecke chain. In particular 
we investigate commuting transfer-matrix type elements which are generating functions for the 
higher Hamiltonians of the model. In Sect. 3 the algebraic description of the fusion procedure for 
the solutions of the reflection equation and for the transfer-matrix type elements are presented. 
In this Section we deduce the functional relations for transfer-matrix type elements of the open 
Hecke chain. In Sections 4 and 5 we discuss the Temperley-Lieb "quotient" of the Hecke model for 
which the closed functional equations for the transfer-matrix type elements can be deduced. We note 
that this analysis is closely related to the approach proposed in \cite{Kuli} for the investigation 
of spin chain systems constructed on the base of the Temperley-Lieb algebra of special $R$- 
matrices. In Conclusion we discuss briefly the interrelations between the algebraic approach of 
this paper and standard reflection equation technique \cite{Skl} developed for the investigation of 
the open chain systems.

\section{Open Hecke chains}
\setcounter{equation}0

A braid group ${\cal B}_{M+1}$ is generated by Artin elements $\sigma_i$ $(i=1, \dots M)$
subject to relations:
\be
\label{braidg}
\sigma_i \, \sigma_{i+1} \, \sigma_i =
\sigma_{i+1} \, \sigma_i \,  \sigma_{i+1} \; , \;\;\;
\sigma_{i} \,  \sigma_{j} = \sigma_{j} \,  \sigma_{i} \;\;
{\rm for} \;\; |i-j| > 1 \; .
\ee

An $A$-Type Hecke algebra $H_{M+1}(q)$ is
a quotient of the group algebra of the braid group ${\cal B}_{M+1}$
by additional Hecke relations
\be
\label{ahecke}
\sigma^2_i - 1 = \lambda \,  \sigma_i \; , \;\; (i = 1, \dots , M) \; ,
\ee
where $\lambda := (q -q^{-1})$ and $q \in {\bf C} \backslash \{0\}$ is a parameter. Let $x \in {\bf 
C}$ be a spectral parameter. We introduce baxterized elements \cite{Jimb1} (see also \cite{Isa}, 
\cite{Isa1})
\be
\lb{baxtH}
\sigma_n(x) : =  \sigma_n - x \sigma_n^{-1} \in H_{M+1}(q) \; ,
\ee
which solve the Yang-Baxter equation
\be
\lb{ybeH}
\sigma_n(x) \, \sigma_{n-1}(xy) \, \sigma_n(y) =
\sigma_{n-1}(y) \, \sigma_n(xy) \, \sigma_{n-1}(x)  \; .
\ee
and satisfy $\sigma_n(x)\sigma_n(y)=\lambda \sigma_n(x y) + (1-x)(1-y)$.
The normalized elements \cite{IsOg1}
\be
\lb{baxtH2}
e^{-}_n (x) :=  \frac{\sigma_n - x \sigma_n^{-1}}{q x - q^{-1}}  \; , \;\;\;
e^{+}_n (x) := \frac{\sigma_n - x \sigma_n^{-1}}{q - q^{-1} x}  \; ,
\ee
obey the unitarity condition
$e_n^{\pm} (x) \, e_n^{\pm} (x^{-1}) =1$.
These baxterized elements are useful for the definition of symmetrizers $A^{+}_{1 \to n}$ and 
antisymmetrizers $A^{-}_{1 \to n}$ ($n=1,2,\dots,M+1$) in the Hecke algebra
$H_{M+1}:=H_{M+1}(q)$.
The operators $A^{\pm}_{1 \to n}$ can be defined
inductively by using recurrent relations \cite{Jimb1} (see also \cite{Isa1} and
references therein)
\be
\lb{asym}
A^{\pm}_{1 \to n+1} = A^{\pm}_{1 \to n} \, e^{\pm}_n (q^{\mp 2n}) \, A^{\pm}_{1 \to n} \; , \;\;\;
A^{\pm}_{1 \to 1} := 1  \; ,
\ee
and obey
\be
\lb{santis1}
\begin{array}{c}
\sigma_i \, A^{\pm}_{1 \to n} = A^{\pm}_{1 \to n} \, \sigma_i =
\pm q^{\pm 1} \, A^{\pm}_{1 \to n} \;\;
(i =1, \dots n-1) \; , \\ \\
A^{\pm}_{1 \to n} \, A^{\pm}_{1 \to m} = A^{\pm}_{1 \to n} =
A^{\pm}_{1 \to m} \, A^{\pm}_{1 \to n} \;\; (\forall n \geq m) \; ,
\;\;\; A^{\pm}_{1 \to n} \, A^{\mp}_{1 \to m} = 0 \; , \\ \\
e^{\pm}_i(x) \, A^{\pm}_{1 \to n} = A^{\pm}_{1 \to n} \, e^{\pm}_i(x) =  A^{\pm}_{1 \to n} \;\;
(i =1, \dots n-1) \; .
\end{array}
\ee

An affine Hecke algebra $\hat{H}_{M+1}$ (see, e.g., \cite{ChPr}, Chapter 12.3)
is an extension of the Hecke algebra $H_{M+1}$. The algebra  $\hat{H}_{M+1}$ is generated
by elements
$\sigma_i$ $(i=1, \dots , M)$
of $H_{M+1}$ and affine generators $y_k$ $(k=1, \dots , M+1)$
which satisfy:
\be
\lb{afheck}
y_{k+1} = \sigma_k \, y_k \, \sigma_k \; , \;\;\; y_k \, y_j = y_j \, y_k \; , \;\;\;
 y_j \, \sigma_i  =  \sigma_i \, y_j \;\; (j \neq i,i+1) \; .
\ee
The elements $\{ y_k \}$ form a commutative subalgebra in $\hat{H}_{M+1}$, while
symmetric functions of $y_k$ form a center in $\hat{H}_{M+1}$.

One can prove by induction that the element
\be
\lb{xxz55}
\begin{array}{c}
y_n(x) =
\sigma_{n-1}(\frac{x}{\xi_{n-1}}) \cdots \sigma_2(\frac{x}{\xi_2}) \sigma_1(\frac{x}{\xi_1}) \, y_1(x) \,
\sigma_{1}(x \xi_1) \sigma_2(x \xi_2) \cdots \sigma_{n-1}(x \xi_{n-1})\; ,
\end{array}
\ee
solves ($\forall \xi_1,\dots,\xi_{n-1} \in {\bf C} \backslash \{ 0 \}$)
the {\it reflection equation}
\be
\lb{reflH}
\sigma_n \left(x \, z^{-1}\right) \, y_n(x) \, \sigma_n(x \, z) \, y_n(z) =
y_n(z) \, \sigma_n(x \, z) \, y_n(x) \,  \sigma_n \left(x \, z^{-1}\right) \; ,
\ee
where
$y_1(x) \in \hat{H}_{M+1}$ is any {\it local}
(i.e., $[y_1(x), \sigma_k]=0$ $\forall k > 1$) solution of (\ref{reflH}) for $n=1$.
In \cite{IsOg1} we have found such local solution
which is rational in $y_1$ function:
\be
\lb{soluH}
y_1(x) = \frac{y_1 - \xi \, x }{y_{1} - \xi \, x^{-1}} \; ,
\ee
where $y_1$ is the affine generator of the algebra $\hat{H}_{M+1}$
and $\xi \in {\bf C}$ is a parameter.
The operator $y_1(x)$ (\ref{soluH}) is regular $y_1(1)=1$, and
obeys the unitarity condition: $y_1(x) y_1(x^{-1}) = 1$.
Below we consider the special form of the element (\ref{xxz55})
(when all $\xi_k =1$)
\be
\lb{xxz5}
y_n(x) = \sigma_{n-1}(x) \cdots \sigma_2(x) \sigma_1(x) \, y_1(x) \,
\sigma_{1}(x) \sigma_2(x) \cdots \sigma_{n-1}(x)\; ,
\ee
which is a Hecke algebra analog of the Sklyanin's monodromy matrix \cite{Skl}.
The elements (\ref{xxz5}) has been used in \cite{IsOg1} for the construction of the integrable
chain systems.

Consider the following inclusions of the subalgebras
$\hat{H}_{1} \subset \hat{H}_{2} \subset \dots \subset \hat{H}_{M+1}$:
$$
\{y_1; \sigma_1, \dots ,\sigma_{n-1}\} = \hat{H}_{n} \subset \hat{H}_{n+1} =
\{y_1; \sigma_1, \dots ,\sigma_{n-1},\sigma_n \} \; .
$$
Then, following \cite{IsOg1} we equip the algebra $\hat{H}_{M+1}$
by linear mappings
$$
Tr_{D(n+1)}: \;\; \hat{H}_{n+1} \to \hat{H}_{n} \; ,
$$
from the algebras $\hat{H}_{n+1}$ to its subalgebras $\hat{H}_{n}$,
such that for all $X,X' \in \hat{H}_{n}$ and $Y \in \hat{H}_{n+1}$ we have
\be
\label{map}
\begin{array}{c}
Tr_{D(n+1)} ( X ) = D^{(0)} \, X  \, , \;\;
Tr_{D(n+1)} ( X \, Y \, X' ) = X \, Tr_{D(n+1)}(Y) \, X' \;\;  \, , \\[0.1cm]
Tr_{D(n+1)} ( \sigma_n^{\pm 1} X \sigma_n^{\mp 1}) =
Tr_{D(n)} (X)  \; , \;\;\;
Tr_{D(n+1)} (X \sigma_n X') =  X \, X' \; , \\[0.1cm]
Tr_{D(1)} (y_1^k)= D^{(k)} \; , \;\; Tr_{D(n)}  Tr_{D(n+1)} ( \sigma_n Y ) =
Tr_{D(n)}  Tr_{D(n+1)} ( Y \sigma_n) \; ,
\end{array}
\ee
where $D^{(k)} \in {\bf C}\backslash \{0\}$ $(k \in {\bf Z})$ are constants.
We stress that $D^{(k)}$ could be considered as additional generators
of an abelian subalgebra $\hat{H}_{0}$ which extends $\hat{H}_{M+1}$
and be central in $\hat{H}_{M+1}$, but for us it is enough to put $D^{(k)}$ to
constants.

Using the maps $Tr_{D(n+1)}$ one can show that the baxterized elements $\sigma_n(x)$ (\ref{baxtH}) 
obey the following identity $(\forall \, X \in \hat{H}_{n}, \; \forall x,z)$:
\be
\lb{mapp1}
Tr_{_{D(n+1)}} \left( \sigma_n(x) \, X \, \sigma_n(z) \right)  =
(1-x) \, (1-z) \,
Tr_{_{D(n)}} (X) + \lambda \left(  1 - \frac{x \, z}{b} \right) \, X \;  ,
\ee
where
$$
b := \frac{1}{1-\lambda D^{(0)}} \; .
$$
Applying this identity to the reflection eq. (\ref{reflH}) we deduce
$$
\left[ \tau_{n-1}(x) , \, y_n(z) \right] =
\frac{\lambda (x b^{-1} - x^{-1})}{\left( (x + x^{-1}) - (z + z^{-1}) \right)  } \,
\left[ y_n(x) , \, y_n(z) \right] \; ,
$$
where operators
\be
\lb{tau11}
\tau_n(x) = Tr_{_{D(n+1)}} \left( y_{n+1}(x) \right)
\ee
form a commutative family $[\tau_n(x), \, \tau_n(z)]=0$ $(\forall x,z)$
(see \cite{IsOg1}).

Now we formulate the open Hecke chains using the operator analogs of
Sklyanin's transfer-matrices \cite{Skl}. In our case these transfer-matrix operators
$\tau_n(x)$ are the
elements of $\hat{H}_{n}$
and represented by (\ref{tau11}) where the solution $y_{n+1}(x)$
of the reflection equation is taken in the special form (\ref{xxz5}):
\be
\lb{tau1}
\tau_n(x) = Tr_{_{D(n+1)}} \left( y_{n+1}(x) \right) =
Tr_{_{D(n+1)}} \left( \sigma_{n}(x) \cdots  \sigma_1(x) \, y_1(x) \,
\sigma_1(x)  \cdots \sigma_{n}(x) \right)  ,
\ee
where $y_1(x)$ is any local and regular solution of (\ref{reflH}) for $n=1$
(e.g. one can take the operator (\ref{soluH})).
The local Hamiltonian of the open Hecke chains is
\be
\lb{ham2}
{\cal H}_n =   \sum_{m=1}^{n-1} \sigma_m - \frac{\lambda}{2} \, y'_1(1)  \; .
\ee
This Hamiltonian (up to a normalization factor and additional constant)
can be obtained by differentiating $\tau_{n}(x)$
with respect to $x$ at the point $x=1$. The Hamiltonian (\ref{ham2})
describes the open chain model with nontrivial boundary condition on the first site
(given by the second term in (\ref{ham2}))
and free boundary condition on the last site of the chain.
The Hamiltonian (\ref{intrH})
is obtained by the substitution of (\ref{soluH}) in (\ref{ham2}).

It follows from (\ref{mapp1}) that
the transfer-matrix operator $\tau_n(x)$ (\ref{tau1}) satisfies recurrent equation
\be
\lb{bethe1}
\tau_n(x) =  \lambda (1 -\frac{x^2}{b}) \, y_{n}(x) +
(1-x)^2 \, \tau_{n-1}(x) \; ,
\ee
which can be solved as
\be
\lb{bethe5}
\tau_n(x) =
\lambda (1 -\frac{x^2}{b})
\left( \sum_{k=0}^{n-1} \, (1-x)^{2k} \, y_{n-k}(x) \right) +
(1-x)^{2n} \, Tr_{_{D(1)}} \left( y_{1}(x) \right) \; ,
\ee
where monodromy elements $y_{k}(x)$ (\ref{xxz5})
are the solutions of the reflection equation (\ref{reflH}) for $n=k$.

Consider the open chain model with free boundary condition
for both sides of the chain, i.e. $y_1(x)=1$. In this case we deduce from (\ref{bethe5})
\be
\lb{triv}
\begin{array}{c}
 \tau_n(x) \left|_{_{y_1(x)=1}}  =
Tr_{_{D(n+1)}} \left( \sigma_{n}(x) \cdots \sigma_2(x) \, \sigma^2_1(x)\,
\sigma_{2}(x) \cdots \sigma_{n}(x) \right) = \right. \\ \\
= \lambda \left(1 -\frac{x^2}{b}\right) J_n(x)
 + (1-x)^{2n} \, D^{(0)} =: T^{(1)}_n(x) \; ,
\end{array}
\ee
where $J_n(x)$ are special polynomials in
spectral parameter $x$ of the order $(2n-2)$:
$$
J_n(x)=
\left( \sum_{k=0}^{n-2} \, (1-x)^{2k} \, \sigma_{n-k-1}(x) \cdots \sigma_1^2(x)
\cdots \sigma_{n-k-1}(x) \right) + (1-x)^{2(n-1)} =
$$
\be
\lb{triv1}
= \lambda \, (1+x) \left( \sum_{a=1}^{2n-3} (1-x)^a \, (\lambda x)^{2n-3-a} \, j_a
\right) + f_n(x) \; ,
\ee
and $f_n(x)$ are scalar polynomial functions of $x$:
$$
f_n(x) = \sum_{k=0}^{n-2} (1-x)^{2k} \left( (1-x)^2 + \lambda^2 x^2 \right)^{n-k-1}
+ (1-x)^{2(n-1)} \; .
$$
The representation (\ref{triv1}) follows from equations
$$
\sigma_k(x) = (1-x) \sigma_k + \lambda x \; , \;\;\;
\sigma^2_k(x) = \lambda (1+x)(1-x) \sigma_k + [(1-x)^2 + \lambda^2 x^2] \; .
$$

Since the transfer - matrix type elements $\tau_n(x)$ form a commutative family
$[\tau_n(x), \tau_n(z)]=0$,
we have from eqs. (\ref{triv}), (\ref{triv1}) a set $\{j_1,j_2, \dots ,j_{2n-3} \}$ of $(2n-3)$ 
commuting elements of the Hecke algebra $H_n$. It is clear that
$$
J_n(0) = \sigma_{n-1} \cdots \sigma^2_{1} \cdots \sigma_{n-1}  +
 \sigma_{n-2} \cdots \sigma^2_{1}  \cdots \sigma_{n-2} + \dots  + \sigma^2_{1} + 1 =
$$
$$
= \lambda \sum_{m=1}^{n-1} \sum_{k=1}^m ( \sigma_{k} \cdots
\sigma_{m-1} \sigma_{m} \sigma_{m-1} \cdots \sigma_{k} ) + n =
\lambda \, j_{2n-3} + f_n(0) \; ,
$$
and the operator $j_{2n-3}$ (as well as $J_n(0)$) is a central element in $H_n$.
On the other hand we have
$$
J_n'(1) = - 2 \lambda^{2n-3}\left(\sum_{m=1}^{n-1} \sigma_m \right)
+ 2 \lambda^{2n-2}(n-1)=
- 2 \lambda^{2n-3} \, j_{1} + f_n'(1) \; ,
$$
and according to (\ref{ham2}) we obtain that
\be
\lb{free}
j_{1} = {\cal H}_n^{(free)} = \sum_{m=1}^{n-1} \sigma_m  \; ,
\ee
is the Hamiltonian for the open Hecke chain with free ends.

For the first nontrivial case $n=3$ we have
the set of commuting elements
\be
\lb{bethe6}
j_1=\sigma_1 + \sigma_2 \; , \;\;\; j_2=\sigma_{1} \sigma_{2} + \sigma_{2} \sigma_{1} \; , \;\;\;
j_3=\sigma_{2}\, \sigma_{1} \, \sigma_{2} + \sigma_1 + \sigma_2 \; .
\ee
The element $j_1$ is the Hamiltonian ${\cal H}_3^{(free)}$ for the open Hecke chain and
$j_3$ is a central element in $H_3$.
The characteristic identity for the Hamiltonian $j_1={\cal H}_3^{(free)}$ is calculated directly:
$$
(j_1 + 2 q^{-1})(j_1 - 2 q)(j_1 - \lambda -1)(j_1 - \lambda +1)=0 \; .
$$
and it means that ${\rm Spec}({\cal H}_3^{(free)}) = (2q,-2q^{-1},\lambda \pm 1)$.
The first two eigenvalues $(\pm 2q^{\pm 1})$
correspond to the one dimensional representations
$\sigma_i = \pm q^{\pm 1}$ $(i=1,2)$ of $H_3$. The doublet
$(\lambda \pm 1)$ corresponds to the 2-dimensional irrep of $H_3(q)$
(for the representation theory of Hecke algebras see
\cite{Isa1}, \cite{IsOgH}, \cite{OgPya} and refs. therein).

For the next case $n=4$ we obtain the following set of commuting elements
$$
\begin{array}{c}
j_1 = \sum_{i=1}^3 \sigma_i  \; , \;\;\;
j_2 = \{ \sigma_1, \, \sigma_2  \}_{+} +
\{ \sigma_2, \, \sigma_3 \}_{+} + 2 \sigma_3 \sigma_1 \; , \\ \\
j_3 =   \{ \sigma_1 \sigma_3 , \, \sigma_2  \}_{+} +
 (\sigma_1 + \sigma_3) \sigma_2  (\sigma_1 + \sigma_3) + \lambda \sigma_3 \sigma_1
+ 2 \sum_{i=1}^3 \sigma_i  \; , \\ \\
j_4 = \{ \sigma_2 \sigma_3  \sigma_2 , \, \sigma_1 \}_{+} +
\{ \sigma_2 \sigma_1  \sigma_2 , \, \sigma_3 \}_{+} +
\{ \sigma_2, \, \sigma_3 \}_{+} +  \{ \sigma_2 , \, \sigma_1 \}_{+}  \; , \\ \\
j_5 = \sigma_1 \sigma_2 \sigma_3  \sigma_2 \sigma_1
+ \sigma_2 \sigma_3  \sigma_2 + \sigma_1 \sigma_2  \sigma_1 + \sigma_1 + \sigma_2 + \sigma_3 \; , \\ \\
\end{array}
$$
The coefficient $j_5$ is a central element in $H_{4}$.
We note that the longest element in $H_4$:
$j= \sigma_1 \sigma_2 \sigma_3  \sigma_1 \sigma_2 \sigma_1=(j_5-j_1)(j_1-\lambda)-j_4$
commutes with the Hamiltonian $j_1={\cal H}_{4}^{(free)}$.
The spectrum of ${\cal H}_{4}^{(free)}$ consists of the eigenvalues:
$(\pm 3 q^{\pm 1})$ for two 1-dimensional irreps, $(2q - q^{-1}, 2q - q^{-1} \pm \sqrt{2})$ and
$(-2q^{-1}+q, -2q^{-1}+q \pm \sqrt{2})$ for two dual 3-dimensional irreps, and
$(\frac{3}{2} \lambda \pm \frac{1}{2} \sqrt{q^{-2}+10 + q^2})$ for 2-dimensional irrep
of $H_4$.

\vspace{0.3cm}
\noindent
{\bf Remark.} Consider the element $j$ of the braid group $B_{M+1}$
$$
j := (\sigma_1  \cdots \sigma_M) (\sigma_1 \cdots \sigma_{M-1}) \cdots
(\sigma_1 \sigma_2) \cdot \sigma_1 \; .
$$
This element commutes with the element
$(\sum_{i=1}^M \sigma_i)$ of the group algebra of $B_{M+1}$, since
$\sigma_i \, j = j \, \sigma_{M+1-i}$. Thus, for the Hecke quotient $H_{M+1}$
of group algebra of $B_{M+1}$,
the element $j \in H_{M+1}$ is a conservation charge
for the model of the chain with the Hamiltonian ${\cal H}_{M+1}^{(free)}$
(\ref{free}). Using the longest element $j \in H_n$
one can show that operators (\ref{triv}), (\ref{triv1})
are invariant under the mirror transformation
$\sigma_k \leftrightarrow \sigma_{n-k}$.
It demonstrates the mirror symmetry of the open Hecke chain described by the
Hamiltonian (\ref{free}) with respect to
the exchange of the first and last sites of the chain.

\section{Fusion for baxterized solutions of reflection \\ equation}

The fusion procedure for the solutions of the Yang-Baxter
equation has been proposed in \cite{KRSk}, \cite{KuSk82}. The fusion for the solutions of the
reflection equation was considered in \cite{MeNe}, where it was
applied to the investigation of the XXZ open spin chain model.
In this Section we formulate the fusion for the solutions
of the reflection equations in terms of the affine Hecke algebra generators.
Below we use concise notations
\be
\lb{not}
e^{x}_n := e^{+}_n(x) \; , \;\;\; y_1^x := y_1(x)
\ee
for baxterized elements $e^{+}_n(x)$ (\ref{baxtH2}) and  solutions
$y_1(x)$ of the reflection equation (\ref{reflH}).

The reflection equation (\ref{reflH}) can be graphically represented  in the form

\unitlength=1cm

\begin{picture}(14,4)

\put(1,1){\line(1,0){5}}

\put(0.5,3){\vector(1,-1){2}}
\put(2.5,1){\vector(1,1){2}}

\put(0.3,3){\vector(3,-2){3}}
\put(3.3,1){\vector(3,2){3}}

\put(2.1,0.7){\tiny $y_n^x$}
\put(3.1,0.7){\tiny $y_n^z$}
\put(2.5,1.6){\tiny $\sigma_n^{x z}$}
\put(1.2,2.5){\tiny $\sigma_n^{x/z}$}

\put(7,1.5){$=$}

\put(8,1){\line(1,0){5}}

\put(9.1,3){\vector(1,-1){2}}
\put(11.1,1){\vector(1,1){2}}

\put(7.3,3){\vector(3,-2){3}}
\put(10.3,1){\vector(3,2){3}}

\put(10,0.7){\tiny $y_n^z$}
\put(11,0.7){\tiny $y_n^x$}
\put(10.6,1.6){\tiny $\sigma_n^{x z}$}
\put(12.1,2.6){\tiny $\sigma_n^{x/z}$}

\put(6,0.2){\bf Fig. 1}

\end{picture}

\noindent
Considering the sequence of the pictures

\unitlength=1cm

\begin{picture}(14,4)

\put(1,1){\line(1,0){5}}

\put(0.5,3){\vector(1,-1){2}}
\put(2.5,1){\vector(1,1){2}}

\put(0.3,3){\vector(3,-2){3}}
\put(3.3,1){\vector(3,2){3}}

\put(2.3,0.67){\tiny $y_1^x$}
\put(3.2,0.6){\tiny $y_1^{x q^2}$}
\put(2.5,1.6){\tiny $\sigma_1^{x^2 q^2}$}
\put(1.2,2.5){\tiny $\sigma_1^{q^{-2}}$}

\put(7,1){\line(1,0){6}}

\put(7.5,3.3){\vector(2,-1){4.5}}
\put(11.9,1){\vector(2,1){3}}

\put(7.5,3.5){\vector(1,-1){2.5}}
\put(10,1){\vector(1,1){2}}

\put(8.1,3.4){\vector(1,-2){1.2}}
\put(9.3,1){\vector(1,2){1}}

\put(9,0.7){\tiny $y_1^x$}
\put(9.9,0.7){\tiny $y_1^{x q^2}$}
\put(11.9,0.7){\tiny $y_1^{x q^4}$}

\put(7.2,2.8){\tiny $\sigma_1^{q^{-2}}$}
\put(8.5,3){\tiny $\sigma_2^{q^{-4}}$}
\put(7.9,2.1){\tiny $\sigma_1^{q^{-2}}$}

\put(9.7,1.4){\tiny $\sigma_1^{x^2 q^2}$}
\put(10,2.1){\tiny $\sigma_2^{x^2 q^4}$}
\put(11,1.7){\tiny $\sigma_1^{x^2 q^6}$}

\put(6,0.2){\bf Fig. 2}

\end{picture}

\noindent
etc., one can easily construct fusions for solutions of the reflection equation. Indeed, the 
algebraic expressions for these diagrams are
\be
\lb{fus12}
\begin{array}{c}
Y_{1\to 2}(x) = e_1^{q^{-2}} \cdot [ y_1^{x} \, e_1^{x^2 q^{2}}
\, y_1^{x q^{2}}] = e_1^{q^{-2}} \cdot y_{1\to 2}(x) \; , \\ \\
Y_{1 \to 3}(x) =[ e_1^{q^{-2}} \, e_2^{q^{-4}} \,
e_1^{q^{-2}}] \cdot
[y_1^{x} \, e_1^{x^2 q^{2}}
\, y_1^{x q^{2}}  \, e_2^{x^2 q^{4}}  \, e_1^{x^2 q^{6}}
\, y_1^{x q^{4}} ] =
[ e_1^{q^{-2}} \, e_2^{q^{-4}} \,
e_1^{q^{-2}}] \cdot y_{1 \to 3}(x) \; ,
\end{array}
\ee
where we have used the concise notations $e^{x}_n,y_1^x$ (\ref{not}).
The formula which generalizes (\ref{fus12}) is
\be
\lb{fre}
Y_{1 \to k}(x) = A^{+}_{1 \to k} \cdot y_{1 \to k}(x) =
\overline{y}_{1 \to k}(x) \cdot A^{+}_{1 \to k} \; ,
\ee
where
$$
\begin{array}{c}
y_{1 \to k}(x) :=
y_1^x \, [e_1^{x^2 q^2}
\, y_1^{x q^2}]  \, [e_2^{x^2 q^4}  \, e_1^{x^2 q^6} \, y_1^{x q^4}]
\, [e_3^{x^2 q^6}  \, e_2^{x^2 q^8}  \, e_1^{x^2 q^{10}} \, y_1^{x q^6}] \cdots \\[0.2cm]
\cdots
[e_{k-1}^{x^2 q^{2(k-1)}} \, e_{k-2}^{x^2 q^{2k}} \,
e_{k-3}^{x^2 q^{2(k+1)}} \cdots
 e_2^{x^2 q^{2(2k-4)}}  \, e_1^{x^2 q^{2(2k-3)}} \, y_1^{x q^{2(k-1)}}] \; ,
\end{array}
$$
and $\overline{y}_{1 \to k}(x)$ is obtained from
$y_{1 \to k}(x)$ by writing it from right to left (the second equality
in (\ref{fre}) is obtained with the help of the Yang-Baxter (\ref{ybeH}) and reflection equations 
(\ref{reflH}) and can be justified immediately from the consideration of its graphical 
representation analogous to Fig. 2). The corresponding fusion for the baxterized elements is given 
by the expression
\be
\lb{fybe}
\begin{array}{c}
\sigma_{1 \to k, k+1 \to 2k}(x) = A^{+}_{1 \to k} \cdot A^{+}_{k+1 \to 2k} \cdot
\sigma_{k+1 \leftarrow 1}^{x} \,
\sigma_{k+2 \leftarrow 2}^{x q^2} \,  \cdots
\sigma_{2k \leftarrow k}^{x q^{2(k-1)}} \equiv \\ \\
\equiv A^{+}_{1 \to k} \cdot A^{+}_{k+1 \to 2k} \cdot \sigma_{k \to 2k}^{x q^{-2(k-1)}} \,
 \cdots \sigma_{2 \to k+2}^{x q^{-2}} \, \sigma_{1 \to k+1}^{x} \; ,
\end{array}
\ee
where
$$
\sigma_{k+1 \leftarrow 1}^{x} = e_k^{x q^{-2(k-1)}}  \cdots
e_{2}^{x q^{-2}} \, e_1^{x} \; , \;\;\;
e_{1 \to k+ 1}^{x} = e_1^{x} e_{2}^{x q^{2}} \cdots e_{k}^{x q^{2(k-1)}} \; .
$$
The elements (\ref{fybe}) solve the highest Yang-Baxter equations.
One can check directly (by using the graphical representation of the type Fig. 2)
that (\ref{fre}) and (\ref{fybe}) satisfy the equation
\be
\lb{frefl}
\begin{array}{c}
\sigma_{1 \to k, k+1 \to 2k}(x \, z^{-1}) \, Y_{1 \to k}(x)  \,
\sigma_{1 \to k, k+1 \to 2k}(x \, z) \, Y_{1 \to k}(z)
= \\ \\
=  Y_{1 \to k}(z)  \, \sigma_{1 \to k, k+1 \to 2k}(x \, z) \,
Y_{1 \to k}(x) \, \sigma_{1 \to k, k+1 \to 2k}(x \, z^{-1}) \; .
\end{array}
\ee
which is a fusion version of the
reflection equation (\ref{reflH}).

The highest transfer-matrix type element
$\tau^{(k)}(x)$ which corresponds to the fusion solution (\ref{fre}) of the
reflection equation (\ref{frefl}) is
\be
\lb{tau3}
\tau^{(k)}(x) = Tr_{_{D(1 \to k)}} \left(
Y_{1 \to k}(x) \, \right) = Tr_{_{D(1 \to k)}} \left( A^{+}_{1 \to k} \cdot
y_{1 \to k}(x) \, \right) \; .
\ee
In particular we have for $k=1,2$
\be
\lb{tau12}
\tau^{(1)}(x) = Tr_{D(1)} \left( y_1^x  \right) \; , \;\;\;
\tau^{(2)}(x) = Tr_{D(1 \to 2)} \left(e_1^{q^{-2}} \, y_1^x \, [e_1^{x^2 q^2}
\, y_1^{x q^2}]  \right) \; .
\ee

Below we need the following Lemma.

\vspace{0.3cm}

\noindent
{\bf Lemma 1.} {\it In the Hecke algebra the following identity holds $(\forall x)$
\be
\lb{lem2}
\begin{array}{c}
A^{+}_{1 \to k} \, [\sigma_{k}^{x}
\sigma_{k-1}^{x q^{2}}
\cdots \sigma_{1}^{x q^{2(k-1)}} ] =
 [\sigma_{k}^{x q^{2(k-1)}}
\sigma_{k-1}^{x q^{2(k-2)}}
\cdots \sigma_{2}^{x q^{2}} \sigma_{1}^{x} ] \, A^{+}_{2 \to k+1}  \; .
\end{array}
\ee
where $A^{+}_{1 \to k}$ are symmetrizers (\ref{asym}).
}

\noindent
{\bf Proof.} We prove eq. (\ref{lem2}) by induction.
Let (\ref{lem2}) is correct for some fixed
$k$. Consider the left hand side of (\ref{lem2}) for
the case $k \to k+1$
$$
\begin{array}{c}
\!\!\!\! A^{+}_{1 \to k+1} \, [\sigma_{k+1}^{x} \sigma_k^{xq^2}
\cdots  \sigma_{1}^{x q^{2k}} ] =
[e_1^{q^{-2}} \cdots e_{k}^{q^{-2k}} ] \,
A^{+}_{1 \to k} \, \sigma_{k+1}^{x} \, [ \sigma_k^{xq^2}
\cdots  \sigma_{1}^{x q^{2k}} ] =
\\ \\
\!\!\!\!
 = (\ref{lem2}) = [e_1^{q^{-2}} \cdots e_{k-1}^{q^{-2(k-1)}}  e_{k}^{q^{-2k}} ]  \,
\sigma_{k+1}^{x} \,
[ \sigma_k^{xq^{2k}} \sigma_{k-1}^{xq^{2(k-1)}} \cdots  \sigma_{1}^{x q^{2}} ] \,
A^{+}_{2 \to k+1} = \\ \\
= (\ref{ybeH}) =
[ \sigma_{k+1}^{xq^{2k}} \sigma_{k}^{xq^{2(k-1)}}\cdots  \sigma_{2}^{x q^{2}} ] \sigma_{1}^{x}
\, [e_2^{q^{-2}} \cdots e_{k}^{q^{-2(k-1)}}  e_{k+1}^{q^{-2k}} ]  \, A^{+}_{2 \to k+1} = \\ \\
= [ \sigma_{k+1}^{xq^{2k}} \cdots  \sigma_{2}^{x q^{2}} \sigma_{1}^{x}]
\, A^{+}_{2 \to k+2}
\end{array}
$$
This equation is equivalent to (\ref{lem2}) for $k \to k+1$. \hfill $\Box$

\vspace{0.5cm}

\noindent
{\bf Proposition 1.} {\it The transfer-matrix type elements $\tau^{(k)}(x)$ (\ref{fre}) satisfy the
following identity
\be
\lb{iden1}
\tau^{(k)}(x)  \tau^{(1)}(x q^{2k}) = \phi_k'(x) \, \tau^{(k+1)}(x)  +
 \phi_k''(x) \, \tau^{(k,1)}(x) \; ,
\ee
where coefficient functions are
$$
\phi_k'(x) = \frac{ \left(1 - x^{2} q^{4k} \, b^{-1} \right)
\left(1 -  x^{2} q^{2(k-1)} \right)}{\left(1 - x^{2} q^{2k} \, b^{-1} \right)
\left(1 -  x^{2} q^{4k-2} \right)} \; ,
$$
$$
\phi_k''(x) = \frac{q^{k} (1-q^{-2k})}{(1-q^{-2(k+1)})(1-q^2)^k} \,
\frac{ \left(1- x^{2} q^{2(k-1)} b^{-1} \right)\left(1 -  x^{2} q^{2(k-1)} \right)}{
\left(1 -  x^{2} q^{2k} b^{-1}\right)
\left(1 -  x^{2} q^{4k-2} \right)} \; ,
$$
and
$$
\begin{array}{c}
\tau^{(k,1)}(x) =
Tr_{_{D(1 \to k+1)}} \left( [\sigma^{q^2}_{1} \sigma_{2}^{q^2}  \cdots \sigma_k^{q^2}
A^{+}_{1 \to k} ] \cdot  y_{1 \to k+1}(x) \right) \; ,
\end{array}
$$
is a new transfer-matrix type element.
}

\vspace{0.1cm}

\noindent
{\bf Proof.}
Using identity (\ref{mapp1}) for $z=b/x$
we deduce
$$
\begin{array}{c}
\eta_k(x) \, \tau^{(1)}(x q^{2k}) =
\eta_k(x) \, Tr_{D(1)} ( y_1^{x q^{2k}} ) = \\ \\
= Tr_{D(k+1)} \left( [e_{k}^{x^2 q^{2k}} \, e_{k-1}^{x^2 q^{2(k+1)}}  \cdots
 e_1^{x^2 q^{2(2k-1)}} ] y_1^{x q^{2k}}
[e_1^{\frac{b}{x^2 q^{2(2k-1)}}}  \cdots e_{k}^{\frac{b}{x^2 q^{2k}}}] \right) \; ,
\end{array}
$$
where
$$
\eta_k(x) = \frac{(1 - x^2 q^{2(2k-1)}) (1 - x^{2} q^{2k} b^{-1})}{
(1 - x^2 q^{2(k-1)}) (1 - x^{2} q^{4k} b^{-1})} \; .
$$
Thus, we have
\be
\lb{prop1}
\begin{array}{c}
\eta_k(x) \, \tau^{(k)}(x)  \tau^{(1)}(x q^{2k}) =
%\\[0.2cm]=
Tr_{_{D(1 \to k+1)}} \left( A^{+}_{1 \to k} \cdot
y_{1 \to k+1}(x)
%[e_{k}^{x^2 q^{2k}}   \cdots e_1^{x^2 q^{4k-2}} \, y_1^{x q^{2k}}] \,
[e_1^{\frac{b}{x^2 q^{4k-2}}}  \cdots e_{k}^{\frac{b}{x^2 q^{2k}}}]
\right)
\end{array}
\ee

One can represent the identity element as linear combination of two baxterized elements
\be
\lb{ident}
1 =  e_k(q^{-2k}) + \frac{(1-q^{-2k})}{(q^2-q^{-2k})\lambda} \,
\sigma_k(q^2) \; , \;\;\;
e_k(q^{-2k}) := \frac{\sigma_k(q^{-2k})}{q-q^{-2k-1}} \; .
\ee
Taking into account eqs. (\ref{ident}), (\ref{lem2}) we write eq. (\ref{prop1}) in
the form
\be
\lb{prop2}
\begin{array}{c}
\eta_k(x) \,
\tau^{(k)}(x)  \, \tau^{(1)}(x q^{2k})
= Tr_{_{D(1 \to k+1)}} \left( (A^{+}_{1 \to k+1} +
\frac{(1-q^{-2k})}{(q^2-q^{-2k})\lambda} \sigma_{k}^{q^2}
A^{+}_{1 \to k} ) \cdot
y_{1 \to k}(x)
 \cdot
\right. \\ \\
\left.
\cdot [e_{k}^{x^2 q^{2k}} \, e_{k-1}^{x^2 q^{2(k+1)}}  \cdots
 e_1^{x^2 q^{4k-2}} y_1^{x q^{2k}}]
e_1^{\frac{b}{x^2 q^{4k-2}}}  \cdots e_{k}^{\frac{b}{x^2 q^{2k}}}
\right)=
\end{array}
\ee
$$
\begin{array}{c}
= (\ref{lem2}) = Tr_{_{D(1 \to k+1)}} \left( (A^{+}_{1 \to k+1} y_{1 \to k+1}(x)
e_1^{\frac{b}{x^2 q^{4k-2}}}  \cdots e_{k}^{\frac{b}{x^2 q^{2k}}} \right) +
\frac{(1-q^{-2k})}{(q^2-q^{-2k})\lambda} \cdot
 \\ \\
 \cdot Tr_{_{D(1 \to k+1)}} \left(
\overline{y}_{1 \to k}(x)
 \cdot
 [e_{k}^{x^2 q^{4k-2}}  \cdots
 e_1^{x^2 q^{2k}} y_1^{x q^{2k}}] A^{+}_{2 \to k+1}
e_1^{\frac{b}{x^2 q^{4k-2}}}  \cdots e_{k}^{\frac{b}{x^2 q^{2k}}} \, \sigma_{k}^{q^2}
\right) \; .
\end{array}
$$
Since $\sigma_{k+1}^{q^{-2}} \, \sigma_{k+1}^{q^{2}}=0$ one can deduce the equation
\be
\lb{sss}
\sigma_{k+1}(q^{-2}) e_{k}(x) \sigma_{k+1}(q^{2}) = \frac{(1-x)}{(1-q^2)(q-q^{-1} x)}
\sigma_{k+1}(q^{-2}) \sigma_{k}(q^{2}) \sigma_{k+1}(q^{2}) \; ,
\ee
which leads to
\be
\lb{ident2}
 A^{+}_{2 \to k+1}
[e_1^{\frac{b}{x^2 q^{4k-2}}}  \cdots  e_{k-1}^{\frac{b}{x^2 q^{2k+2}}} e_{k}^{\frac{b}{x^2 q^{2k}}}
\sigma^{q^2}_{k} ] =
 \frac{(-q^{-k-1}) (1-\frac{b}{x^2 q^{2(k-1)}}  )}{(1-q^2)^{k-1}(1-\frac{b}{x^2 q^{4k}}  )}
 A^{+}_{2 \to k+1} [\sigma_1^{q^2}  \cdots  \sigma^{q^2}_{k} ] \; .
\ee
We have also the following identity (see last eqs. in (\ref{santis1}))
\be
\lb{iden3}
\begin{array}{c}
A^{+}_{1 \to k+1} \, [e_1^{\frac{b}{x^2 q^{4k-2}}}  \cdots e_{k}^{\frac{b}{x^2 q^{2k}}}] =
A^{+}_{1 \to k+1} \; .
\end{array}
\ee

Finally using (\ref{ident2}) and (\ref{iden3}) we obtain for (\ref{prop2})
$$
\begin{array}{c}
\eta_k(x) \,
\tau^{(k)}(x)  \tau^{(1)}(x q^{2k}) =  \tau^{(k+1)}(x)  +  \\ \\
\!\!\!\!\!\!\!\!\!\!\!\!
+  \frac{q^{-k-2} (1-q^{-2k})}{(1-q^{-2(k+1)}) (1-q^2)^{k}} \,
 \frac{ (1-\frac{b}{x^2 q^{2(k-1)}}  )}{(1-\frac{b}{x^2 q^{4k}}  )}
 \, Tr_{_{D(1 \to k+1)}} \left(
A^{+}_{1 \to k}  \cdot  y_{1 \to k+1}(x) \,
[\sigma_1^{q^2}  \cdots  \sigma_{k-1}^{q^2} \sigma^{q^2}_{k} ]
\right) \; ,
\end{array}
$$
which is equivalent to (\ref{iden1}). \hfill $\Box$

\section{Temperley-Lieb quotient for the Hecke chain model.}

Eqs. (\ref{iden1}) are not closed with respect to the set of
the transfer-matrix type elements $\tau^{(k)}(x)$ (\ref{tau3}).
To obtain closed set of relations we
need to consider special quotients of the Hecke algebra $H_{M+1}$.
In this Section we consider a Temperley-Lieb quotient $H^{(TL)}_{M+1}$
which is defined by imposing the additional constraints
on the generators $\sigma_i \in H_{M+1}$ $(i=1, \dots ,M)$:
\be
\lb{tlq}
\sigma_i(q^2) \sigma_{i \pm 1}(q^4) \sigma_i(q^2) = 0  \; .
\ee
These constraints are equivalent to the
vanishing of the third rank antisymmetrizers $A^{-}_{i \to i+2} = 0$ (\ref{asym}). Moreover, they
fix the parameter $D^{(0)}$ (or $b$):
\be
\lb{bbb}
\begin{array}{c}
0 = Tr_{_{D(i+1)}} \left( \sigma_{i-1}(q^2) \sigma_{i}(q^4) \sigma_{i-1}(q^2) \right) =
\sigma_{i-1}(q^2) \left( Tr_{_{D(i+1)}} \sigma_{i}(q^4)\right) \sigma_{i-1}(q^2) = \\[0.2cm]
= \sigma_{i-1}^2(q^2)(1 - q^4(1- \lambda D^{(0)})) = \sigma_{i-1}^2(q^2)(1 - q^4/b) \;\;
\Rightarrow \\[0.2cm]
\Rightarrow  \;\; b = q^4 \; , \;\;\; D^{(0)} = \frac{1-q^{-4}}{\lambda} \; .
\end{array}
\ee

A Temperley-Lieb (TL) quotient of the affine Hecke algebra, in addition to
(\ref{tlq}), is defined by imposing the constraint on the affine generator
of $\hat{H}_{M+1}$:
\be
\lb{tlhh}
y_1 \, \sigma_1 \, y_1 \, \sigma_1^{q^2} = \sigma_1^{q^2} \, y_1 \, \sigma_1 \, y_1  =
\Delta \, \sigma_1^{q^2} \; ,
\ee
where $\Delta = \frac{\lambda q}{(1-q^{-4})} (D^{(2)}- q (D^{(1)})^2) \in \hat{H}_{0}$
is a central element
in $\hat{H}_{M+1}$. In the presence of the
map (\ref{map}), eq. (\ref{tlhh}) requires that the affine generator $y_1$
obeys quadratic characteristic identity:
$$
y_1^2 - q \, D^{(1)} \, y_1 = q^{-1} \, \Delta \; ,
$$
and after that
(\ref{tlhh}) is written in the form
$\sigma_1^{q^2} \, y_1 \, \sigma_1^{q^2} = q^2(q^2-1)D^{(1)} \sigma_1^{q^2}$.
Thus, this factor of the affine Hecke algebra coincides with the one-boundary
TL, or blob, algebra \cite{Lev}, \cite{MS}.

The constraint (\ref{tlhh}) also leads to the constraints
\be
\lb{fus6}
 y_1^{z q^{-2}} \, e_1^{z^2 q^{-2}} \, y_1^{z} \, \sigma_1^{q^2} =
 \sigma_1^{q^2} \, y_1^{z} \, e_1^{z^2 q^{-2}} \, y_1^{z q^{-2}} =:
\Delta^{(0)}(z)\, \sigma_1^{q^2} \; ,
\ee
where $y_1^{z} \in \hat{H}_{M+1}$ is any local solution
of the reflection equation (\ref{reflH}) for $n=1$
and $\Delta^{(0)}(z)$ is a central element in $\hat{H}_{M+1}$.

\vspace{0.2cm}

\noindent
{\bf Proposition 2.} {\it In the case of the TL quotient for
the open Hecke chain, the functional relations (\ref{iden1}) are closed
with respect to the family of the transfer-matrix type elements $\tau^{(k)}(x)$:
\be
\lb{prop4}
\begin{array}{c}
\tau^{(k)}(x)  \tau^{(1)}(x q^{2k}) =  \phi_k'(x) \, \tau^{(k+1)}(x)  +
\phi_k'''(x)  \, \Delta^{(0)}(xq^{2k}) \, \tau^{(k-1)}(x)
\end{array}
\ee
where $\tau^{(0)}(x) :=1$ and
$$
\phi_k'(x) := \frac{\left(1 -  x^{2} q^{2(k-1)} \right)
\left(1 - x^{2} q^{4k-4}  \right)}{\left(1 - x^{2} q^{2k-4}  \right)
\left(1 -  x^{2} q^{4k-2} \right)} \; ,
$$
$$
\phi_k'''(x) := \frac{1}{(-q^2)}
\frac{ \left(1- x^{2} q^{2k-6} \right)\left(1 -  x^{2} q^{4k-4} \right)}{
\left(1 -  x^{2} q^{2k-4} \right)
\left(1 -  x^{2} q^{4k-2} \right)} \; .
$$
}

\noindent
{\bf Proof.}
We write (\ref{iden1}) in the form
\be
\lb{prop3}
\begin{array}{c}
\tau^{(k)}(x)  \tau^{(1)}(x q^{2k}) = \phi_k'(x) \, \tau^{(k+1)}(x)  +
 \phi_k''(x) \, \Delta^{(0)}(xq^{2k}) \cdot \\ \\
\!\!\!\!\!\! \cdot Tr_{_{D(1 \to k+1)}} \left(
A^{+}_{1 \to k}
\cdot y_{1 \to k-1}(x)   \cdot
[e_{k-1}^{x^2 q^{2(k-1)}}  \cdots
 e_1^{x^2 q^{4k-6}}]
[e_{k}^{x^2 q^{2k}}  \cdots
 e_2^{x^2 q^{4(k-1)}}]
[\sigma_1^{q^2}  \cdots  \sigma^{q^2}_{k} ]
\right) \; .
\end{array}
\ee
Constraints (\ref{tlq}),
which defines the TL quotient for the Hecke algebra, lead to
the following identities
$$
\begin{array}{c}
\sigma_i(q^2) \, \sigma_{i \pm 1}(x) \, \sigma_i(q^2) = \lambda q^3 (1-q^{-4} x) \, \sigma_i(q^2) \; , \\ \\
e_i(x) \, e_{i \pm 1}(x q^2) \, \sigma_i(q^2) = - \xi(x) \, \sigma_{i\pm 1}(q^2) \, \sigma_i(q^2)
\;\;\;\; (\forall x) \; ,
\end{array}
$$
where $\xi(x) = \frac{(1-x q^2)}{q^2 \lambda (1-x)}$.
We use these identities to simplify (\ref{prop3})
by means of the following relation
$$
\begin{array}{c}
[e_{k-1}^{x^2 q^{2(k-1)}}  \cdots e_1^{x^2 q^{4k-6}}]
[e_{k}^{x^2 q^{2k}}  \cdots e_2^{x^2 q^{4(k-1)}}]
[\sigma_1^{q^2}  \cdots  \sigma_{k-1}^{q^2} \sigma^{q^2}_{k} ] = \\ \\
= [e_{k-1}^{x^2 q^{2(k-1)}}  \cdots e_2^{x^2 q^{4k-8}}]
[e_{k}^{x^2 q^{2k}}  \cdots e_3^{x^2 q^{4k-6}}]
e_1^{x^2 q^{4k-6}} e_2^{x^2 q^{4k-4}} \sigma_1^{q^2}
[\sigma_2^{q^2}  \cdots  \sigma_{k-1}^{q^2} \sigma^{q^2}_{k} ] = \\ \\
\!\!\!\!\!\!
(-\xi(x^2 q^{4k-6})) \, [e_{k-1}^{x^2 q^{2(k-1)}}  \cdots e_2^{x^2 q^{4k-8}}]
[e_{k}^{x^2 q^{2k}}  \cdots e_3^{x^2 q^{4k-6}}]
\sigma_2^{q^{2}} \sigma_1^{q^2}
[\sigma_2^{q^2}  \cdots  \sigma_{k-1}^{q^2} \sigma^{q^2}_{k} ] = \dots = \\ \\
= \prod_{n=1}^{k-1}
(-\xi(x^2 q^{4k-4-2n})) \,  [\sigma_{k}^{q^{2}}  \sigma_{k-1}^{q^2} \cdots
\sigma_2^{q^{2}} \sigma_1^{q^2}
\sigma_2^{q^2}  \cdots  \sigma_{k-1}^{q^2} \sigma^{q^2}_{k} ] = \\ \\
= (\lambda q)^{2(k-1)} \, \prod_{n=1}^{k-1}
(-\xi(x^2 q^{4k-4-2n})) \,   \sigma^{q^2}_{k} = (-\lambda)^{k-1} \,
\frac{1 - x^2 q^{4(k-1)}}{1 - x^2 q^{2(k-1)}} \,  \sigma^{q^2}_{k} \; .
\end{array}
$$
Taking into account this relation and identities
$$
Tr_{_{D(k+1)}} (\sigma^{q^2}_{k}) = 1-q^2/b \; , \;\;\;
 Tr_{_{D(k)}} A^{+}_{1 \to k}  = \frac{1-\frac{q^{-2(k-1)}}{b}}{q-q^{-2k+1}} \, A^{+}_{1 \to k-1}
$$
which follow from the definitions (\ref{asym}), (\ref{map}),
we write (\ref{prop3}) for $b=q^4$ in the form (\ref{prop4}). \hfill $\Box$

 Eqs. (\ref{prop4}) are closed with respect to
the set of commuting elements $\tau^{(k)}(x)$. We note that eqs. (\ref{prop4})
can be used for definition of elements $\tau^{(k)}(x)$ in the case
when integers $k <0$.

\section{$T-Q$ Baxter equation for open chains of TL type}

Now we make a change of variables $z =x q^{2k}$ in (\ref{prop4})
and put
$$
Q^{(k)}(z):=(1-z^2 q^{-2(k+2)})\, \tau^{(k)}(z q^{-2k}) \; .
$$
As a result we write (\ref{prop4}) in the form
\be
\lb{fus5}
 \frac{(1 - z^2 q^{-2}) }{(1- z^2 q^{-4})} \,  Q^{(k)}(z) \,  \tau^{(1)}(z)
=  Q^{(k+1)}(z q^2)
- \frac{\Delta^{(0)}(z)}{q^{2}}
\, Q^{(k-1)}(z q^{-2}) \; ,
\ee
where $\tau^{(1)}(z)$ and $\Delta^{(0)}(z)$ are defined in (\ref{tau12}) and
(\ref{fus6}), respectively. Introduce the generating function $Q(z,w)$
of the elements $Q^{(k)}(z)$
$$
Q(z,w) := \sum_{k = - \infty}^\infty \, w^k \, Q^{(k)}(z) \; .
$$
Then eq. (\ref{fus5}) is represented in the form
of the $T-Q$ Baxter equation
\be
\lb{tqb}
 \frac{(1 - z^2 q^{-2}) }{(1- z^2 q^{-4})} \,  Q(z,w) \,  \tau^{(1)}(z)
=  \frac{1}{w} Q (z q^2,w)
- \frac{\Delta^{(0)}(z)  \, w}{q^{2}}
\, Q (z q^{-2},w) \; .
\ee

We remind that our aim is to find the spectrum of the
transfer-matrix type operator $\tau^{(1)}(z)$
for some fixed determinant function $\Delta^{(0)}(z)$.
We fix the explicit expression for the determinant $\Delta^{(0)}(z)$ (\ref{fus6})
by considering the special model of open chain with $N$ sites.
In this case the definition of $\Delta^{(0)}(z)=:\Delta^{(0)}_N(z)$ is
\be
\lb{fus7}
\Delta^{(0)}_{N}(z)\, \sigma_{N+1}^{q^2}=
 y_{N+1}^{z q^{-2}} \, e_{N+1}^{z^2 q^{-2}} \, y_{N+1}^{z} \, \sigma_{N+1}^{q^2}
\ee
where monodromy element $y_{N+1}^{z}$ is a normalized version of (\ref{xxz5}) and given by
\be
\lb{fus10}
y_{N+1}^{z} = e^+_{N}(z)  y_{N}^z e^+_{N}(z)= \dots
= e^+_{N}(z) \cdots e^+_1(z) \, y_1^z \,
e^+_1(z) \cdots e^+_{N}(z) \; .
\ee
Let $y_1^z$ be a polynomial in $z$ up to
a multiplication by some inessential scalar function of $z$.
Such solutions $y_1^z$ of the reflection equation have been proposed in
\cite{Levi}, \cite{MudK}.
We note that for the case when $y_1^z$ is a polynomial in $z$ of order $K$
the transfer-matrix type element is represented in the form
\be
\lb{fus11}
\tau^{(1)}_{N}(z) = Tr_{_{D(N+1)}} (y_{N+1}^{z}) = \frac{1}{(q-zq^{-1})^{2N}} \tau_{N}(z) \; ,
\ee
where the element $\tau_{N}(z)$ (\ref{bethe5}) is a polynomial in $z$ of order $2N +K$.
Then, from (\ref{fus7}) we have
$$
\Delta^{(0)}_{N}(z)\, \sigma_{N+1}^{q^2}=
(e_N^{z q^{-2}} y_N^{z q^{-2}} e_N^{z q^{-2}}) \,
e_{N+1}^{z^2 q^{-2}} \, (e_N^{z} y_N^{z} e_N^{z}) \, \sigma_{N+1}^{q^2} =
$$
$$
= e_{N}^{z q^{-2}}  e_{{N+1}}^{z} y_{N}^{z q^{-2}} \,
e_{N}^{z^2 q^{-2}} \, y_{N}^{z} \, (e_{{N+1}}^{z q^{-2}} \, e_{N}^{z} \, \sigma_{N+1}^{q^2}) =
$$
$$
\!\!\!\!
= -\xi(z/q^{2}) \, e_{N}^{z q^{-2}}  e_{{N+1}}^{z} (y_{N}^{z q^{-2}}
e_{N}^{z^2 q^{-2}}  y_{N}^{z} \, \sigma_{N}^{q^2})  \sigma_{N+1}^{q^2} =
(\xi(z/q^{2}))^2 \sigma_{{N+1}}^{q^2} [\Delta^{(0)}_{N-1}(z)  \sigma_{N}^{q^2}] \,
\sigma_{N+1}^{q^2} =
$$
$$
= \dots = (\xi(z q^{-2}))^{2 N} \, \sigma_{{N+1}}^{q^2} \cdots \sigma_{2}^{q^2}
\, [y_1^{z} \, e_1^{z^2 q^{-2}} \, y_1^{z q^{-2}}
\sigma_{1}^{q^2}] \, \sigma_{2}^{q^2} \cdots \sigma_{N+1}^{q^2} =
$$
$$
=\Delta^{(0)}_{0}(z) \left( \xi(z q^{-2}) \lambda q \right)^{2 N} \, \sigma_{N+1}^{q^2}
$$
or
\be
\lb{fus8}
\Delta^{(0)}_{N}(z) = \Delta^{(0)}_{0}(z) \left( \xi(z q^{-2}) \lambda q \right)^{2 N} =
\Delta^{(0)}_{0}(z) \left( \frac{(1-z)}{(q-zq^{-1})}  \right)^{2 N} \; .
\ee
We choose the trivial
boundary condition $y_1(z)=1$ which corresponds
to the model of the open chain with free ends described by the Hamiltonian
${\cal H}_{N}^{(free)}$
(\ref{free}). In this case the initial determinant $\Delta^{(0)}_{0}(z)$ is fixed by
$$
\Delta^{(0)}_{0}(z) \, \sigma_{1}^{q^2} =
e_1^{z^2 q^{-2}} \,  \sigma_{1}^{q^2} = -q^{-2} \frac{(1-z^2)}{(1-z^2 q^{-4})}
\, \sigma_{1}^{q^2} \; .
$$
Thus, the functional equations (\ref{fus5}) are represented as
\be
\lb{fus55}
\begin{array}{c}
\frac{(1 - z^2 q^{-2}) }{(1- z^2 q^{-4})} \,  Q^{(k)}(z) \,
 \frac{T^{(1)}_N(z)}{(q-zq^{-1})^{2 N}}
%= \\ \\
=  Q^{(k+1)}(z q^2)
+ q^{-4} \frac{(1-z^2)}{(1-z^2 q^{-4})} \left( \frac{(1-z)}{(q-zq^{-1})}  \right)^{2 N}
\, Q^{(k-1)}(z q^{-2})
\end{array}
\ee
where we substitute
$\tau^{(1)}_N(z) = \frac{T^{(1)}_N(z)}{(q-zq^{-1})^{2 N}}$
and $T^{(1)}_N(z)$ is the polynomial in $z$ of order $2N$ (\ref{triv})
which has been considered in Sect. 2.

Eqs. (\ref{tqb}) and (\ref{fus55}) could be used
(by applying the standard procedure) for
derivation of analytical Bethe ansatz equations which contains
the information about the spectrum of the
transfer-matrix type operators $\tau^{(1)}(z)$ and, in particular,
about the spectrum of the Hamiltonians (\ref{ham2}).

\section{Conclusion}

In Conclusion we briefly discuss the interrelations between the algebraic approach presented above 
and standard technique developed for the investigation of the open spin chain systems and based on 
the matrix reflection equation \cite{Skl}.

Consider the $R$-matrix representation $\rho_R$ of the affine Hecke algebra $\hat{H}_{M+1}$:
\be
\lb{rhoR}
\begin{array}{c}
\rho_R(\sigma_i) = \hat{R}_{i i+1} := I^{\otimes (i-1)} \otimes \hat{R} \otimes I^{\otimes (M-i)}   \; ,
\\ \\
\rho_R(y_1) = S(L^{-}_1) \,  L^{+}_1 := (1/L^{-}) \,  L^{+} \otimes I^{\otimes M}  \; .
\end{array}
\ee
Here $I$ is $(n+m) \times (n+m)$ identity matrix;
$\hat{R} \in {\rm End}(V_{n+m}^{\otimes 2})$ is the
fundamental $R$-matrix for $U_q(gl(n|m))$ written in the braid form \cite{Isa},
\cite{Isa1} (for the standard form see \cite{DKS}):
$$
\hat{R}=\sum_i (-1)^{[i]} \, q^{1 - 2[i]} \, e_{ii} \otimes e_{ii} +
 \sum_{i \neq j} (-1)^{[i][j]}  \, e_{ij} \otimes e_{ji} +
\lambda \,  \sum_{j > i} \, e_{ii} \otimes e_{jj} \; ,
$$
$$
\hat{R}^2 =\lambda \hat{R} + 1 \; , \;\;\;
(-1)^{^{[1][2]}} \, \hat{R}_{12} = \hat{R}_{12} \, (-1)^{^{[1][2]}} \; , \;\;
\lim_{q \to 1} \left( \hat{R}_{12} \right) = {\cal P}_{12} \; ,
$$
where $\lambda = q-q^{-1}$, $e_{ij}$ are matrix units, $[i] = 0,1 (mod(2))$
denotes the parity of the components of supervectors in $V_{n+m}$,
${\cal P}_{12} := (-1)^{^{[1][2]}} P_{12}$ is a superpermutation matrix
($P_{12}$ is a permutation matrix)
and we have used concise matrix notations, e.g.,
$((-1)^{^{[1][2]}})^{i_1 i_2}_{j_1 j_2} =
(-1)^{[i_1][i_2]} \delta^{i_1}_{j_1} \delta^{i_2}_{j_2}$.
The operator-valued matrices $L^{+}(L^{-}) \in {\rm Mat}(n+m)$
are invertible, upper (lower) triangular, and satisfy:
\be
\lb{frt}
\begin{array}{c}
\hat{R}_{12} \,  L^{\pm}_2 \, (-1)^{^{[1][2]}} \, L^{\pm}_1 =
L^{\pm}_2 \, (-1)^{^{[1][2]}} \, L^{\pm}_1  \, \hat{R}_{12}  \; , \\[0.2cm]
\hat{R}_{12} \,  L^{+}_2 \,(-1)^{^{[1][2]}} \, L^{-}_1 =
L^{-}_2 \, (-1)^{^{[1][2]}} \, L^{+}_1 \,  \hat{R}_{12} \; .
\end{array}
\ee
We prescribe to the elements $(L^{\pm})^i_j$ the grading $([i]+[j])$.
According to the approach of \cite{FRT},
$L^{\pm}$ are matrices of Cartan type generators of $U_q(gl(n|m))$.
Note that, in the representation $\rho_R$ (\ref{rhoR}), the solution (\ref{soluH}) has the form
\be
\lb{rhoy}
\rho_R(y_1(x)) =
\frac{1}{(L^+_1 - \xi x^{-1} L^-_1)}(L^+_1 - \xi x L^-_1) =: K_1(x)\; ,
\ee
and one can immediately check that (\ref{rhoy}) solves the reflection equation
(\ref{reflH}) written in the $R$-matrix form
\be
\lb{reflHr}
\hat{R}_{12} (x/z) \, K_1(x) \,
\hat{R}_{12}  (x \, z) \, K_1(z) =
K_1(z)\, \hat{R}_{12} (x \, z) \, K_1(x) \,
\hat{R}_{12}  (x/z) \; ,
\ee
where
$$
\hat{R}_{n n+1} (x):= \hat{R}_{n n+1} - x \hat{R}_{n n+1}^{-1} = \rho_{R}(\sigma_n) \; ,
$$
is the $R$-matrix image of the baxterized element (\ref{baxtH}). To prove (\ref{reflHr}) one needs 
only the fact that $K(x)$ (\ref{rhoy}) is represented in the factorized form $L^{-1}(\xi/x) L(\xi 
x)$, where the operator-valued matrix
\be
\lb{eval}
L(x) = (L^+ - x L^-) \; ,
\ee
defines the evaluation representation \cite{ChPr}
of the affine superalgebra $U_q(\hat{gl}(n|m))$ and
satisfies the intertwining relations
\be
\lb{intertw}
\hat{R}_{12}(x) \, L_2(x y) \, (-1)^{^{[1][2]}} \, L_1(y) =
L_2(y) \, (-1)^{^{[1][2]}} \, L_1(x y) \, \hat{R}_{12}(x) \; .
\ee
We note that all grading factors $(-1)^{^{[1][2]}}$ have disappeared in (\ref{reflHr}).
The comultiplications for the
quantum superalgebras with defining relations (\ref{frt}) and (\ref{intertw})
have the usual form $\Delta L^{\pm} = L^{\pm} \otimes L^{\pm}$
and $\Delta L(x) = L(x) \otimes L(x)$, where $\otimes$
is a graded tensor product, i.e., $(a \otimes b)(c \otimes d)=(-1)^{^{[b][c]}}(a c \otimes b d)$.

We note that reflection equation
solutions of the type (\ref{rhoy})
were called in \cite{Slav} as solutions which admit a "regular factorization".
Given a matrix $L(x)$ which satisfies (\ref{intertw}) one can construct
by means of the standard quantum inverse scatering method an
operator-valued monodromy matrix of the chain with $(N+1)$ sites
\be
\lb{calT}
{\cal T}_{a}(\xi,\xi_1, \dots,\xi_N | x) =
L_a(\xi x) \otimes L_a(\xi_1 x) \otimes \dots \otimes L_a(\xi_{N} x) \; ,
\ee
where $a$ denotes the number of auxiliary matrix space
and $\otimes$ is a graded tensor product of the operator spaces.
The monodromy matrix (\ref{calT}) satisfies the intertwinig relations (\ref{intertw})
$$
\hat{R}_{12}(x) \,
{\cal T}_2(\xi, \dots |x y) \, (-1)^{^{[1][2]}} \, {\cal T}_1(\xi, \dots|y) =
{\cal T}_2(\xi, \dots|y) \, (-1)^{^{[1][2]}} \, {\cal T}_1(\xi, \dots|x y) \,
\hat{R}_{12}(x) \; .
$$
Then, we construct the two-row Sklyanin's monodromy matrix
\be
\lb{sklmm}
{\cal K}_a(\xi,\xi_1, \dots,\xi_N | x) = {\cal T}^{-1}_{a}(\xi,\xi_1, \dots,\xi_N | x^{-1}) \cdot
{\cal T}_{a}(\xi,\xi_1, \dots,\xi_N | x) \; ,
\ee
which obviously solves the reflection equation (\ref{reflHr}).
The corresponding Sklyanin's transfer matrix is given by the quantum supertrace
\be
\lb{strma}
\tau_N(\xi,\xi_1, \dots,\xi_N | x) = Tr_a(D_a \, {\cal K}_a(\xi,\xi_1, \dots,\xi_N | x))  \; ,
\ee
where
$D^i_j = (-1)^{^{[i]}} \, q^{2m + (-1)^{^{[i]}}(2i-2m-1)} \, \delta^i_j$ \cite{Isa1}
is the matrix of quantum supertrace (for the case of quantum superalgebra $U_q(gl(n|m))$
and one can check that this matrix is a constant solution of the
conjugated reflection equation. Taking into account the fact that
the monodromy matrix (\ref{calT})
satisfies the intertwining relations (\ref{intertw}) one can use a
matrix representation in which the image of (\ref{calT}) is
$$
\begin{array}{c}
{\cal T}_{a}(\xi,\xi_1, \dots,\xi_N | x)  \, \rightarrow \,
(-1)^{^{([1]+\dots +[N])[a]}}
L_a(\xi x) P_{1a} \hat{R}_{1a}(\xi_1 x) \cdots P_{_{Na}} \hat{R}_{_{N a}}(\xi_{_N} x)
= \\ \\
= {\cal P}_{_{Na}} {\cal P}_{_{N-1 N}} \dots  {\cal P}_{23} {\cal P}_{12}
L_1(\xi x) \hat{R}_{12}(\xi_1 x) \dots
\hat{R}_{_{N-1 N}}(\xi_{_{N-1}} x) \hat{R}_{_{N a}}(\xi_{_N} x) \; ,
\end{array}
$$
where we have applied the matrix homomorphism to all factors in (\ref{calT})
except the first one and $1,2, \dots, N$ denote the numbers of matrix spaces
which replace the operator spaces.
Then, for the Sklyanin's monodromy matrix (\ref{sklmm}) we obtain
the representation
$$
\begin{array}{l}
{\cal K}_a(\xi,\xi_1, \dots,\xi_N | x) \, \rightarrow \,
\\ \\
\hat{R}^{-1}_{_{N a}}(\frac{\xi_N}{x}) \hat{R}^{-1}_{_{N-1 N}}(\frac{\xi_{N-1}}{x}) \dots
\hat{R}^{-1}_{12}(\frac{\xi_1}{x}) \frac{1}{L_1(\xi/x)} L_1(\xi x) \hat{R}_{12}(\xi_1 x) \dots
\hat{R}_{_{N-1 N}}(\xi_{_{N-1}} x) \hat{R}_{_{N a}}(\xi_{_N} x) =
\end{array}
$$
\be
\lb{sklmm1}
\begin{array}{c}
= \rho_R \left( e^{+}_{N}(\frac{x}{\xi_N}) e^{+}_{N-1}(\frac{x}{\xi_{N-1}}) \dots
 e^{+}_{1}(\frac{x}{\xi_1}) \, \frac{y_1-\xi x}{y_1-\xi/x} \,  e^{+}_{1}(\xi_1 x) \dots
 e^{+}_{N-1}(\xi_{_{N-1}} x)  e^{+}_{N}(\xi_{_N} x) \right)
\end{array}
\ee
where we have taken into account the identity
$\hat{R}(x^{-1})\hat{R}(x)= (q  - x q^{-1})(q- x^{-1}q^{-1})$,
put $a=(N+1)$ and used the notations
(\ref{baxtH2}), (\ref{rhoR}). It is clear that (\ref{sklmm1}) is the
$R$-matrix image of (\ref{fus10}) for $x=z$, $\xi_i=1$ and
$y_1^x = \frac{y_1-\xi x}{y_1-\xi/x}$.
The quantum trace (\ref{strma}) of (\ref{sklmm1}) coincides with the $R$-matrix image
$\rho_R$ of the transfer matrix type element (\ref{fus11}).

Thus, we have demonstrated that the integrable models
based on the consideration of the monodromy matrices (\ref{sklmm})
and models based on (\ref{fus10})
in general are different and coincides only on their $R$ matrix
projections. This identification is clarified by the formula (\ref{sklmm1}).

\vspace*{-2pt}

\section*{Acknowledgments}
I thank P.P. Kulish for useful
 discussions and comments. I also
thank A.F. Os'kin, S.Z. Pakuliak and P.N. Pyatov for the
discussions of the results of Section 2.


\begin{thebibliography}{99}

\bibitem{IsOg1} A.P. Isaev and O.V. Ogievetsky,
On Baxterized Solutions of Reflection Equation and Integrable Chain Models, 
 Nucl.\ Phys.\  B {\bf 760} (2007) 167; [arXiv:math-ph/0510078].
  %%CITATION = NUPHA,B760,167;%%


\bibitem{Levi} D. Levi and P. Martin, Hecke algebra solutions
to the reflection equation, J. Phys. A {\bf 27} (1994) L521-L526.

\bibitem{MudK} P.P.Kulish and A.I.Mudrov,
Baxterization of solutions to reflection equation with Hecke R-matrix, Lett. Math. Phys. 75
(2006) 151; [arXiv:math.QA/0508289].

\bibitem{Jimb1} M.~Jimbo,
A q-analogue of $U_q(gl(N+1))$, Hecke algebra and the Yang-Baxter
equation, Lett.\ Math.\ Phys.\  {\bf 11} (1986) 247.

\bibitem{FRT} L.D. Faddeev, N.Yu. Reshetikhin, and L.A. Takhtajan,
Quantization Of Lie Groups And Lie Algebras,
Lengingrad Math.\ J.\  1 (1990) 193.


\bibitem{Isa}
A.P. Isaev, Quantum groups and Yang-Baxter equations,
Sov.\ J.\ Part.\ Nucl.\  {\bf 26} (1995) 501
(Fiz. Elem. Chastits i At. Yadra {\bf 26} (1995) 1204).
%%CITATION = SJPNA,26,501;%%

\bibitem{KRSk} P.P. Kulish, N.Yu. Reshetikhin and E.K. Sklyanin, 
 Yang-Baxter equation and representation theory: I,
 Lett. Math. Phys. 5 (1981) 393.

\bibitem{KuSk82} P.P. Kulish and E.K. Sklyanin,
Quantum spectral transform method.
Recent developments, in ``Integrable Quantum Field Theories'',
Lect. Notes in Phys. {\bf 151} (1982) 61.
%%CITATION = LNPHA,151,61;%%

\bibitem{MeNe} L. Mezinchescu and R. Nepomechie, Fusion procedure for
open chains, J. Phys. A: Math. Gen. 25 (1992) 2533.


\bibitem{Zho} Y.K. Zhou,
Row Transfer Matrix Functional Relations for Baxter's
Eight-Vertex and Six-Vertex Models with Open Boundaries Via
More General Reflection Matrices, Nucl. Phys. {\bf B 458} (1996) 504,
hep-th/9510095.


 \bibitem{NZ} W.~L.~Yang, R.~I.~Nepomechie and Y.~Z.~Zhang,
Q-operator and T-Q relation from the fusion hierarchy,
  Phys.\ Lett.\ B {\bf 633} (2006) 664, hep-th/0511134.
%%CITATION = HEP-TH 0511134;%%

\bibitem{Kuli} P.P. Kulish, On spin systems related to the Temperley-Lieb algebra,
J. Phys. A: Math. Gen. {\bf 36} (2003) L489.

\bibitem{Skl} E.K. Sklyanin,
Boundary Conditions For Integrable Quantum Systems,
J.\ Phys.\ A 21 (1988) 2375.
%%CITATION = JPAGB,A21,2375;%%

\bibitem{ChPr} V. Chari and A. Pressley,
A guide to quantum groups, Cambrige Univ. Press (1994).

\bibitem{Isa1} A.P. Isaev, Quantum groups and Yang-Baxter equations,
preprint MPIM (Bonn), MPI 2004-132 (2004), \\
(http://www.mpim-bonn.mpg.de/html/preprints/preprints.html).

\bibitem{IsOgH} A.P. Isaev and O.V. Ogievetsky, On representations of Hecke algebras,
Czechoslovak Jour. of Physics, Vol.55, No.11 (2005) 1433; Representations of A-type Hecke algebras, 
arXiv:0912.3701 [math.QA]. 

\bibitem{OgPya} O.V.Ogievetsky and P.N.Pyatov, Lecture on Hecke algebras,
preprint MPIM (Bonn), MPI 2001-40, \\
(http://www.mpim-bonn.mpg.de/html/preprints/preprints.html).

\bibitem{Lev} D. Levy, The algebraic structure of translation
    invariant spin 1/2 XXZ and Q Potts quantum chains.
Phys. Rev. Lett., {\bf 67} (1991) 1971;
Lattice algebras and the hidden symmetry of
the 2-d Ising model in a magnetic field.
Int. J. Mod. Phys. {\bf A6} (1991) 5127.

\bibitem{MS} P.P. Martin and H. Saleur, The blob algebra and the
periodic Temperley-Lieb algebra. Let. Math. Phys. {\bf 30} (1994) 189,
hep-th/9302094; \\
P.P. Martin and D. Woodcock, On the Structure of the Blob Algebra,
Journal of Algebra, {\bf 225} No. 2 (2000) 957.


\bibitem{DKS}  E.V. Damaskinsky, P.P. Kulish, and M.A. Sokolov,
 Gauss Decomposition for Quantum Groups and Supergroups,
 Zap. Nauch. Semin. POMI 211 (1995) 11-45, q-alg/9505001.


\bibitem{Slav} H. Frahm and N. Slavnov,
New solutions to the reflection equation and
the projecting method, J. Phys. A {\bf 32} (1999) 1547.



\end{thebibliography}
\end{document}